\documentclass[12pt]{article}
\RequirePackage{graphicx}
\usepackage{amsmath, amssymb}
\usepackage{slashed}
\usepackage{type1cm}
\usepackage{amsfonts}
\usepackage{cite}
\usepackage{xcolor}
\usepackage{bm} 
\usepackage[legacycolonsymbols]{mathtools} 

\begin{document}
\sloppy
\vskip 1.5 truecm
\centerline{\large{\bf Quantum field theory on curved manifolds}}
\vskip .75 truecm
\centerline{\bf Tomohiro Matsuda
\footnote{matsuda@sit.ac.jp}}
\vskip .4 truecm
\centerline {\it Laboratory of Physics, Saitama Institute of
 Technology,}
\centerline {\it Fusaiji, Okabe-machi, Saitama 369-0293, 
Japan}
\vskip 1. truecm
\makeatletter
\@addtoreset{equation}{section}
\def\theequation{\thesection.\arabic{equation}}
\makeatother
\vskip 1. truecm
\begin{abstract}
\hspace*{\parindent}
This paper discusses how particle production from the vacuum can be
 explained by local analysis when the field theory is defined by
 differential geometry on curved manifolds.
We have performed the local analysis in a mathematically rigorous way,
 respecting the Markov property.
The exact WKB is used as a tool for extracting non-perturbative effect
 from the local system.
After a serious application of the differential geometry and the exact
 WKB to particle production, we show that entanglement
 does not appear in the 
Unruh effect as far as the standard formulation by the differential geometry
 is valid.
This result should not be attributed to a consistency problem between
 the ``entanglement state'' and the ``standard field
 theory by differential geometry'', but to the fact that the
 conventional calculation of the Unruh effect is done by extrapolation
which is not consistent with the differential geometry.
The situation is similar to that of the Dirac monopole, but topology is
 not relevant and the basis for building field theories in differential
 geometry is strongly involved.
\end{abstract}

\newpage
\section{Introduction}
\hspace*{\parindent}
The knowledge gained by constructing field theory by differential
geometry is significant.
The most famous is what happned in 70's to the solution of the Dirac
monopole\cite{Dirac:1931kp,Wu:1975es}, where the differential geometry
revealed that at least two solutions are involved for the Dirac monopole. 
The solution is famous in topology, but the mathematician's deep
understanding of local analysis in differential 
geometry is at the root of this story.
Similar discoveries are likely to continue in the future, as differential
geometry can involve various mathematical concepts to construct the
field theory in a variety of sophisticated
ways.
In this paper, the Schwinger effect\cite{Schwinger:1951nm} and the Unruh
effect\cite{Unruh:1976db,Hawking:1975vcx,Birrell:textbook} are analysed 
using the standard formulation of the field theory in terms of
differential geometry. 
We paid particular attention to the Markov property and taken great
care to ensure that the computation is localised and consistent with
differential geometry.

When the curvature of a manifold is given by a non-zero constant, a
dynamical element of particle generation may be added to the system, even
though the manifold itself seems to be static.
The Schwinger effect\cite{Schwinger:1951nm} and Hawking
radiation\cite{Hawking:1975vcx} are typical examples of such phenomena.
Particle production in spacetime with curvature has been treated by
various methods\cite{Birrell:textbook}, but we are dissatisfied with
conventional calculations because of loss of locality.
In this paper, we focus on the most primitive method for calculating
Bogoliubov transformations: analysis with field equations. 
In this case, the Bogoliubov transformations are realized as the Stokes
phenomena of the differential equation.
The main reason for us to stick to differential equations is that they
are compatible with differential geometry.
In the study of particle production, when it is analyzed using field
equations, it is customary to introduce asymptotic
states (flat space at infinity in time or space) to define the
``vacuum'' and the particle number there.
One might argue that the loss of locality is inevitable in such
calculation, but at the same time, it is hard to believe that loss of
locality is an essential feature of the phenomena.
Indeed, if particle production is to be a heat bath, the loss of
Markov property seems to be fatal for the analysis.
In principle, it is also strange that the calculations cannot be
completed at the place where the particles are created, without
using the distant past or the distant future.
Therefore, the loss of locality is clearly due to ``for convenience'',
which is the problem with the
computational procedures of the physics side.
Turning to mathematics, gauge theory and general relativity are
constructed by differential geometry and manifolds, where local
consistency is quite rigorous\cite{Kobayashi-Nomizu:textbook, Nakahara:textbook,
Hamilton:textbook}.
It is therefore natural to ask whether, using mathematical concepts,
these problems can be solved without assuming 
asymptotic states at a distance, or without introducing extrapolation.
It could be difficult to notice the unnaturalness we are discussing
about when one is accustomed to ordinary calculations in physics.
However, we strongly insist that loss of locality is
a serious problem that should be resolved by 
mathematics.
In particular, in mathematical concepts of manifolds, tangent spaces
appear naturally. 
These tangent spaces have
properties quite similar to the asymptotic states but appear in local.
To explain in more detail, an operation called local trivialization is
required at the point of contact with the tangent space, and this
operation ensures that the connection is always zero at that point.
The most familiar tangent space in physics will be the local inertial
system, which is of course important in physics.\footnote{
To be more precise, the ``inertial system'' mentioned above is the
tangent space with the observer's (subjective) inertial frame.
(The ``frame'' in the differential geometry is defined on the
tangent space.)
In physics, it is known that the subjective inertial frame causes Thomas
precession\cite{Misner:textbook}.}
The reasons why a tangent space can be used as a vacuum are explained in
detail in this paper, considering a simple example where the vacuum {\it must}
be defined in the {\it local} tangent space.
Indeed, considering how to introduce Lorentz symmetry into the local
system, it is 
obvious that a vacuum must be defined in the tangent space even if it
is defined as an asymptotic state at a distance.
This point will be clarified in the discussion of charts and frames.
Since both Lorentz and gauge symmetries are treated equivalently in the
framework of differential geometry, it is quite natural to assume that
the tangent space to the principal bundle of a gauge transformation has
the same physical meaning.

Note that the particle production discussed here is not an ``overtly
dynamical phenomenon'' like prereheating after inflation, but rather a 
``phenomenon that causes dynamical particle production despite its apparent
static state''.
We will explain later that for the latter case a non-trivial treatment
regarding gauge and Lorentz symmetries is required, which is difficult
to be
described without the help of differential geometry.
In this paper, we will try to explain mathematical ideas as intuitive as
possible, but since the concepts of the mathematics are rather complex,
at least a basic knowledge of a typical differential geometry course is
required to understanding the calculations correctly.
Unlike differential geometry, which could be familiar as an undergraduate
mathematics course, it would be very difficult to understand the exact
WKB without reading any of the references.
The serious difficulty with the exact WKB is that it looks so much like a
regular WKB approximation that it is sometimes very hard to imagine 
how different the mathematics behind it is.

The aim of this paper is to rethink locality of particle production,
keeping the discussions as faithful as possible to mathematics.
The main obstacle to the local description of non-perturbative particle
production has been to find a local analysis of non-perturbative effect,
which is now solved by the exact WKB in mathematics.
The best-known tangent space in physics would be the local inertial system of
general relativity.
In the electromagnetic theory, it might be strange to define a
tangent space, but indeed the tangent space of the principle bundle is
commonly used in mathematics to define the curvature of the
electromagnetic field. 
Our perspective is that particle production on manifolds, such as the
Schwinger and the Unruh effects should be
discussed in a unified way by using the same basic concepts of the
differential geometry and manifolds.
The crucial difference between the Unruh effect\cite{Unruh:1976db} and
the Schwinger effect will be made clear in this paper, by using local
calculations faithful to mathematics.

When attempting to solve locally what has long been solved
globally in physics, it is necessary to reconsider various
mathematical properties that have previously been ignored.
One might wonder why such mathematical concepts are
required even in situations where solutions by path integrals are easily given.
On this point, it would be important to emphasize here that path integral and
Feynman diagrams are so cleverly constructed that a user can sometimes
get the right result without 
having to worry about such mathematical concepts.
If the computation by path integrals is locally complete, our analysis
will not lead to any new conclusions.

Section \ref{sec-diff} of this paper describes the basic concepts of
differential geometry and of manifolds, and the introduction of the
symbols that are used in this paper.
Since differential geometry has almost never been seriously discussed in
the analysis of the Unruh effect, we will be particularly careful to
explain which aspects we will focus on.
As the concept of manifolds is not explained from scratch, the reader
who is not familiar with the mathematical approach is advised to read
the references.

Section \ref{sec-Sch} describes the Schwinger effect when the electric field is
constant or time-dependent.
For fermions, it looks like the Landau-Zener transition\cite{Zener:1932ws}.
By considering the case where the electric field is weakly depending on time,
we show why the vacuum {\it must be} defined in the {\it local} tangent space.
This is the simplest example that clearly demonstrates that finding a naive
global exact solution to an equation cannot always be a correct
answer.
Note that the basic concept explaining that a single solution cannot
cover the whole space is the same as that of the Dirac monopole,
although it has nothing to do with topology.
In order to understand the causes of this non-trivial situation, it is
necessary to take a closer look at the treatment of gauge and Lorentz
symmetries in differential geometry, especially at the necessity of
local trivialization.
At the time of the Dirac monopole, much attention was paid to
topology, but in the root of the discussions there has been 
the basic procedures in finding {\it local} solutions in differential
geometry.

Section 4 discusses a local analysis of the Unruh effect and Hawking
radiation.
Unlike the Unruh effect, where entangled pair production is expected at distant
wedges, Hawking radiation is a pair production localised at the horizon.
Therefore, we can calculate local particle production of Hawking
radiation rigorously using 
differential geometry and the exact WKB, without indication of any
discrepancy between the standard result.

\section{An introduction to differential geometry and manifolds for
 particle production}
\label{sec-diff}

First, we briefly describe the basic concepts of differential
geometry and manifolds for the field theory.
For our argument in this paper, a clear distinction between ``charts''
and ``frames'' is most important.
Therefore, the main purpose of this introduction is to explain clearly
why they are distinguished in mathematics and what are the implications
for physics.

We know that in physics the vacuum is exactly the same for observers in
any frame. 
In mathematics, the Lie algebra and bundles are used to define the situation.
However, the Lie algebra and the bundles are not defined for ``charts''. 
As the Rindler coordinate system is a ``chart'', it was (at least in
principle) unnatural to define a ``vacuum'' in the Rindler chart to discuss
the Bogoliubov transformation.
Furthermore, the ``moving frame'' of an observer has its applicable
range.
The range is estimated as $\propto 1/a$ in Ref.\cite{Misner:textbook},
where $a$ is the rate of acceleration.
The extrapolation beyond this range is not rigorous, but such
extrapolation is normally used in
the conventional calculation of the Unruh effect. 
Therefore, there is a need for a method that allows rigorous
calculations to be made in local.
Our idea is to introduce the exact WKB for the local analysis, using the
vierbeins. 
Without understanding the unnaturalness described
here, it will be hard to understand the following arguments.

Although we will try to provide an overview as intuitive as possible, a
basic knowledge of ordinary differential geometry is essential for a
proper understanding of the subject.
The standard description used here can be found in
Ref.\cite{Kobayashi-Nomizu:textbook, Nakahara:textbook,
Hamilton:textbook}, but in principle different (and more complicated)
extensions of manifolds could be possible in the description of the field
theory\cite{Penrose:1967twistor}.
The mathematical concepts associated with field theory are not
always confined to the basic scenario presented here.

\begin{figure}[ht]
\centering
\includegraphics[width=0.8\columnwidth]{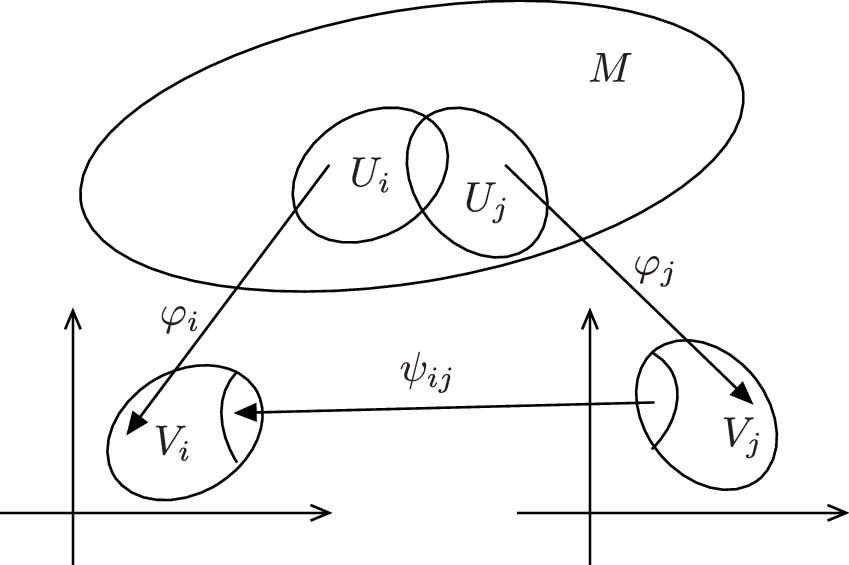}
 \caption{A local coordinate function $\varphi_i$ is shown as a
 homeomorphism from $U_i\in M$ to $V_i\in \mathbb{R}^n$. A chart is a
 pair $(U_i, \varphi_1)$.
The Rindler coordinate is a typical chart on flat spacetime. The
 ``chart'' must be discriminated from the ``frame''.} 
\label{fig_chart}
\end{figure}

A manifold is a fundamental concept in mathematics.
It is defined as a topological space that locally resembles Euclidean
space.
This means that for every point in the manifold, there exists a
neighborhood that can be mapped homeomorphically (i.e., through a
continuous, bijective function with a continuous inverse) to an open
subset of $\mathbb{R}^n$, where $n$ is the dimension of the manifold.
For physics, this implies that in principle the manifold will always have
the required structure that is needed to define a local vacuum.\footnote{To
understand this more clearly for the Schwinger effect, we need to
rethink tangent spaces after introducing gauge symmetries and matter
fields.}  
The local structure of the differential geometry was designed from purely mathematical
considerations.
We will list here the most important definitions (and notations) of the
manifolds that are convenient for our later discussions.
In field theory, it is {\it not} possible to explain everything with just one
manifold.
Such complex structure is important for introducing gauge and Lorentz
symmetries.
\begin{enumerate}
\item Local Euclidean Property: A manifold $M$ is an $n$-dimensional
      manifold if, for every point $p\in M$, there exists an open
      neighborhood $U_i$ of $p$ such that there is a homeomorphism
      $\varphi_i$ : $U_i\rightarrow V_i$, where $V_i$ is an open subset of
      $\mathbb{R}^n$. 
      Note that $\bigcup_i U_i=M$.
      $\varphi_i(p)=(x^1,x^2,...,x^n)$ is called a local
      coordinate of $p$ or simply a coordinate of $p$. $\varphi_i$ for
      $U_i$ is called the coordinate function of $U_i$. 
\item Charts and Atlases: 
 \begin{itemize}
  \item A chart is a pair $(U_i,\varphi_i)$. (See fig.\ref{fig_chart}.)
  \item An atlas is a collection of charts that covers the entire
	manifold, allowing for transitions between different charts
	through coordinate transformations (using the coordinate functions).
\end{itemize}
\end{enumerate}

\begin{figure}[ht]
\centering
\includegraphics[width=0.8\columnwidth]{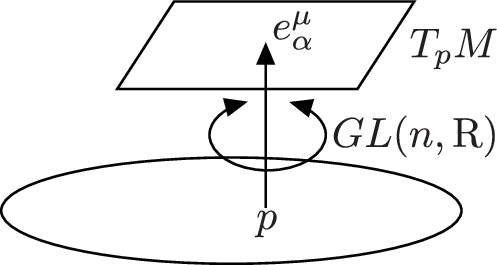}
 \caption{The tangent space $T_pM$ is spanned by
 $\{e_\mu\}=\{\partial/\partial x^\mu\}$, where $\{x^\mu\}$ is the
 coordinate function of $U_i$. As is shown in Eq.(\ref{eq-lie}) and 
the above picture, the vierbeins are essential for defining the Lie algebra.
This means that {\it the vacuum must be defined in the tangent space} as far as 
the vacuum respects the Lorentz symmetry, even if it is placed at far
 distance as to describe the asymptotic state.
The frame of an accelerating observer is called a moving frame, as it
 (the vierbeins) is changed by the Lorentz transformation time to time.} 
\label{fig-tangent}
\end{figure}

Then, to describe the tangent space of a manifold, we define a tangent
vector and a tangent space as follows:
\begin{itemize}
 \item A tangent vector: We introduce a tangent vector at $p$ as
\begin{eqnarray}
X&=&\sum_\mu X^\mu \frac{\partial}{\partial x^\mu}.
\end{eqnarray}
\item A tangent space: We introduce a tangent space as the space
      constructed by the whole tangent vectors at $p$. For concreteness,
      when the bases are given by $\{\frac{\partial}{\partial x^1},
      ...,\frac{\partial}{\partial x^n}\}$, we have 
\begin{eqnarray}
T_p M&=&\left\{ X^\mu \left.\frac{\partial}{\partial x^\mu} \right| X^\mu\in
	 \mathbb{R}\right\},
\end{eqnarray}
where $T_pM\simeq \mathbb{R}^n$.
\end{itemize}
Then, a tangent bundle is defined by\footnote{The tangent bundle is also
a $2n$-dimensional manifold.}
\begin{eqnarray}
TM &\coloneqq&\bigcup_{p\in M}T_p M.
\end{eqnarray}
In this context, a vector field is a projection given by
\begin{eqnarray}
X \colon M &\rightarrow& TM \nonumber\\
p&\rightarrow& X(p)\in T_pM.
\end{eqnarray}
Suppose that $\varphi_i(p)$ is the coordinate function $\{x^\mu (p)\}$
of $U_i$.
In the ``coordinate basis'', $T_p M$ is spanned by
$\{e_\mu\}=\{\partial/\partial x^\mu\}$, while the ``non-coordinate
bases'' is explained as
\begin{eqnarray}
\label{eq-lie}
\hat{e}_\alpha&=&e_\alpha^\mu\frac{\partial}{\partial x^\mu},\,\,\,\,
e_\alpha^\mu\in GL(m,\mathbb{R}),
\end{eqnarray}
where the coefficients $e_\alpha^\mu$ are called vierbeins.
Note that the Lie brackets can be introduced only for the non-coordinate
bases where the vierbeins play an essential role, as is shown in
Fig.\ref{fig-tangent}.
The figure shows that the vacuum must be defined in the tangent space,
even if it is defined as an asymptotic state.

Also, the cotangent space at the point $p$, denoted as $T^*_pM$, is
defined as the dual space of the tangent space $T_pM$.
In this paper, both the tangent space and the cotangent space can simply
be referred to as the tangent space if there is no possibility of
confusion.
For the sake of the explanation that follows, we are going to start with
an intuitive explanation about defining the vacuum in the tangent space.
We are now going to construct a field theory on $M$ as
manifolds\footnote{We need many kinds of manifolds at the same time to
explain the basic properties of the field theory.}: the
field theory constructed on $M$ is, of course, also defined at $p\in M$.
Then, the theory defined on $p$ is naturally extended into the tangent space.
A key feature of the theory extended in the tangent space is that curvatures
vanish.
This is trivial with respect to the spacetime curvature, as the tangent
space is obviously flat, but it should not be trivial with respect to
gauge field curvature, as the gauge field curvature may not vanish in
the flat spacetime.
The vanishing curvature of the gauge field is understood if the gauge
theory in the tangent space (of the spacetime) is also described by the
tangent space of the principal bundle of the gauge.
Thus, by treating both Lorentz and gauge symmetries equally, the local
vacuum can be defined as having no curvature in either sense.
Since a tangent space can be defined on any space defined as a
manifold, here the vacuum is defined using these tangent spaces.
What will be important in our calculations is that the connection is
always zero at the contact point with the tangent space, where 
the local vacuum is defined.

Above, we have described tangent vector bundles to introduce the
concept of bundles.
It seems obvious that one can consider similar
vector bundles in general (not necessarily for tangent vectors of $M$).
This is the idea behind the fibre bundle.
The coordinate functions defined in tangent vector bundles naturally had
a direct product structure.\footnote{Intuitively, the local
direct product structure gives a point where one can start the
discussion with a global symmetry (because they are given by the direct
product) when describing a local symmetry.
This procedure is one that we often see in field theory,
whether or not there is a strict definition of local trivialization.}
When considering general fibre bundles, the direct product structure has
to be introduced by hand.
To clarify the notation used, we refer to the definitions below.
\begin{itemize}
\item Fibre bundle: A fibre bundle is defined by $(E,\pi, M, F, G)$,
      where
 \begin{enumerate}
  \item $E$ is the total space
  \item $M$ is the base space
  \item The projection $\pi$: $E\rightarrow M$ is a continuous
	surjection known as the projection map. The inverse
	$F\coloneqq\pi^{-1}(p)$ is called the fibre at $p\in M$.
  \item The structure group $G$ acts on $F$ from the left.
  \item The local trivialization is introduced by $\phi_i$ for the open
	coverings $\{U_i\}$ of $M$ as (sometimes $\phi_i^{-1}$ instead
	of $\phi_i$ is called the local trivialization because
	$\phi_i^{-1}$ seems to realize the direct product structure.)
\begin{eqnarray}
\phi_i \colon U_i \times F&\rightarrow& \pi^{-1}(U_i)\nonumber\\
(p,f)&\rightarrow& \phi_i(p,f),
\end{eqnarray}
where $E$ locally becomes a direct product.
\end{enumerate}
\end{itemize}
Thanks to the introduction of a local direct product structure in the
fibre, connections can be defined in a natural way.
At this point, the introduction of a direct product structure may not
seem to make sense as physics.
However, when one considers the role that
local inertial systems actually play in general relativity, one has to
consider that it does make sense as physics.
Since in mathematics, general relativity and gauge theories are described
in a unified way, it seems to be natural to treat the two in a unified
way.
Here, the ``gauge field with curvature'' yields the Schwinger effect,
while the ``spacetime with curvature'' yields the Unruh effect and Hawking
radiation. 
We believe that essential differences are highlighted only when
analyses are held on the same foundations as far as possible.

 Before introducing the connection, we should mention a scalar field
 used in the field theory of physics, which appears as a section of $E$
 of a manifold.
A ``section'' of a manifold will need some explanation.
If we think of the elementary function $y=f(x)$ as a map
 $\mathbb{R}\rightarrow\mathbb{R}$, this function cuts $\mathbb{R}$ at
 the destination and gives a single value; if we draw 
 a graph of $y=f(x)$ on the $xy$ space, its appearance will look like
 considering a section of the $xy$ space.
Literally, a scalar field of the field theory corresponds to a section
 of the corresponding mapping. 
More abstractly, suppose that we are given some kind of mapping;
If we give this mapping a concrete form, we are looking at a section.
For cases where the mapping has internal degrees of freedom, a
 section is seen as one concrete form is selected.
In physics, selecting the frame of a particular observer in special
 relativity gives a section of the frame bundle.
For an accelerating observer, the observer's frame moves on the frame
 bundle.
This non-trivial section is called the moving frame.
The range of applicability of the moving frame has been calculated in
Ref.\cite{Misner:textbook}.
It is only consistent in the range $\propto a^{-1}$, where $a$ is the
 accelerating rate of the observer.
This point is very important, since the conventional calculation
 extrapolate well beyond this range and this may have led to an
 entanglement that should not have occurred.
Hypotheses of this kind can only be tested by means of local
 calculations without extrapolation.

It is particularly important not to confuse conventional ``gauge
fixing'' with such a ``section''.
To understand the essence, recall the difference between the metric
$g_{\mu\nu}$ in the theory of gravity and the gauge fixing of its
fluctuation $\delta g_{\mu\nu}$.
We will later analyze particle production in the vicinity of the contact
point of a tangent space, but be
careful not to confuse ``the section used to define the
contact point of the tangent space'' with ``gauge fixing of quantum
fluctuations''. 

Above, fibres have been introduced locally by means of local
trivialization, and the fibres are laminated together to form a fibre bundle.
When local trivialization varied from place to place, it was
necessary to connect them by using the transformation.
Therefore, it is very natural to ask whether
the fibre bundle that is obtained in this simple way really has
the correct differential structure.
If one actually prepares a vector bundle\footnote{The vector bundle has
any dimension and is not always supposed to be the tangent vector bundle.} 
and its section (e.g, a scalar
field) and simply takes the derivative, it can be seen that the simple
derivative is not covariant.
Then, connections are introduced to solve this problem.
This is called a covariant derivative.
This connection is necessary for the differential geometry because fibre
bundles are locally trivialized and then laminated together using
transformations. 
If trivialization is possible in global, then inevitably the
connections will disappear.\footnote{On the other hand, if the section
is defined for a moving frame, the connections do not vanish for the
observer, even if the fibres are globally trivial on the manifold.
The same is true for a ``moving gauge'' of the Schwinger effect.
This point will be explained in more detail later. }
For the same reason, the value of the connection is supposed to vanish at the
contact point of the tangent space.
The reason why non-trivial connections appear in the equations of an
observer in accelerated motion in flat space is very easy to understand
from a differential geometry point of view.
This is because, although the space-time itself is flat, the observer's
equation sees a non-trivial section of the bundle.

To explain the situation using a concrete example, we consider a two-dimensional real vector field $\phi$:
$M\rightarrow \mathbb{R}^2$ as a section of a vector bundle $E$,
where $M=\mathbb{R}^3$ and $F=\mathbb{R}^2$.
Since the fibre is a two-dimensional real space, we choose the structure
group as $GL(2,\mathbb{R})$.
Therefore, $\phi$ translates as $\phi'(x)=g(x)\phi(x)$ by $g(x)\in
GL(2,\mathbb{R})$ and laminated on the fibre.
The question is if $d\phi(x)$ could also be a section of the bundle
on which $GL(2,\mathbb{R})$ can act properly to be laminated on the fibre as the
original function $\phi$.\footnote{The exterior derivative is used
here for simplicity of notation.}
The process of deriving the connection one-form ($A$) is the same as in field
theory and is therefore omitted.
What is important here is that by using 
\begin{eqnarray}
\nabla \phi &=&(d+A)\phi,
\end{eqnarray}
where 
\begin{eqnarray}
A'&=&-(dg)g^{-1}+gAg^{-1},
\end{eqnarray}
one will see 
\begin{eqnarray}
(\nabla \phi)'&=&g\nabla \phi,
\end{eqnarray}
which translates the same way as the original section (i.e, the scalar
field $\phi$) of the vector bundle.
A simple explanation is that since the two sections $\phi$ and $\nabla
\phi$ transform in the same way, they can be laminated in the same way.
This is the requirement for the consistency of the differential geometry.
Here $A$ is called a connection or a gauge field.
Although the transformation of the scalar field is trivial under the
coordinate transformation, the effect of gravity is incorporated in a
natural way as $d\phi$ is also a section of the frame bundle.
The frame bundle will be described later.
 
Given that Lorentz and gauge symmetries are treated as
equivalence classes in field theory, one might think that the above
discussion does not 
adequately describe the situation.
To describe this point, it is necessary to introduce a
principal bundle.
The covariant derivative defined above can also be explained using the
tangent space of the principal bundle.
To give an overview without using further mathematical definitions, consider the
simplest tangent space for an example.
In the tangent space, there are various ways of taking coordinates depending
on the coordinate transformation, so the principal bundle is the fibre
that brings them all together.
The principal bundle deals with such equivalence as a fibre where 
the coordinate transformation induces a motion on it.
Note that in physics, the presence of an observer naturally defines a
section of the principal bundle (frame bundle), since the
observer chooses a unique frame.
The importance of such a frame in physics (called a moving
frame\cite{Misner:textbook} for an accelerating observer) has
already been confirmed by the Thomas precession\cite{Misner:textbook} in
a non-trivial way.
In addition to the principal bundle, a spinor bundle is required if fermions are to be
introduced\cite{Hamilton:textbook}.
However, further explanation is beyond the scope of this paper. 
The reader is referred to the relevant
textbooks\cite{Kobayashi-Nomizu:textbook, Nakahara:textbook,
Hamilton:textbook} for more details.

\section{The Schwinger effect on manifolds}
\label{sec-Sch}
First, we consider the case where the curvature of the manifold is
defined for a gauge symmetry.
The simplest model uses the electromagnetic $U(1)$ gauge symmetry on a
flat space-time and the curvature is introduced by a constant electric field.
In this model, the manifold is static in the sense that the curvature is
constant, but quantum theory expects a dynamical phenomenon (particle
production) on it.
The particle production in this model is called the Schwinger
effect\cite{Schwinger:1951nm}.
In this case, the quickest way to avoid tedious discussions about
manifolds is to use a powerful computational tool, the path
integral\cite{Schwinger:1951nm, Affleck:1981ag, Gelis:2015}.
However, in this paper, we venture a primitive analysis based on field
equations to look closely at what happens in differential geometry and manifolds.
The analysis on the manifold using a scalar field is already given in
Ref.\cite{Enomoto:2022mti, Matsuda:2023mzr, Matsuda:2024ydu}.
The analysis of the fermionic Schwinger
effect as the Landau-Zener transition and its application to a
time-dependent electric field is new. 
Using this model and a slowly varying electric field, we show why 
the concept of the manifold and definition of {\it local} vacuum in the
tangent space is particularly important.

To understand the fermionic Schwinger effect as the Landau-Zener
transition, we introduce the conventional decomposition of the Dirac fermion
as\cite{Peloso:2000hy, Enomoto:2021hfv, Enomoto:2022nuj}
\begin{eqnarray}
\psi&=&\int\frac{d^3k}{(2\pi)^3}
e^{-i\boldsymbol k\cdot \boldsymbol x}\sum_s\left[
u_{\boldsymbol k,s}(t)a_{\boldsymbol k,s}
+v_{\boldsymbol k,s}(t)b^{\dagger}_{-\boldsymbol k,s}\right],
\end{eqnarray}
where $v_{\boldsymbol k,s}=C\left(
\bar{u}_{\boldsymbol k,s}
\right)^T$, and $\psi$ obeys the single-field Dirac equation
\begin{eqnarray}
(i\hbar\slashed{\partial}-m)\psi&=&0.
\end{eqnarray}
Taking the momentum ${\boldsymbol k}\equiv k_z$ and
defining\footnote{Hereafter, we omit ${\boldsymbol k}$ in the indices of $u$.}
\begin{eqnarray}
u_s&\equiv&\left[\frac{u_+}{\sqrt{2}}\psi_s,
			    \frac{u_-}{\sqrt{2}}\psi_s\right]^T,\nonumber\\
v_s&\equiv&\left[\frac{v_+}{\sqrt{2}}\psi_s,
			    \frac{v_-}{\sqrt{2}}\psi_s\right]^T,
\end{eqnarray}
where $\psi_+\equiv (1,0)^T$ and $\psi_-\equiv (0,1)^T$ are eigenvectors
of the helicity operator.
Carefully following the formalism given in Ref.\cite{Peloso:2000hy}, 
one will find
\begin{eqnarray}
\hbar\dot{u}_\pm&=&ik u_{\mp}\mp i m u_\pm,
\end{eqnarray}
which can be written in the matrix form as
\begin{eqnarray}
i\hbar\frac{d}{dt}\left(
\begin{array}{c}
u_+\\
u_-
\end{array}
\right)&=&\left(
\begin{array}{cc}
m& -k \\
 -k & -m
\end{array}
\right)
\left(
\begin{array}{c}
u_+\\
u_-
\end{array}
\right).
\end{eqnarray}
Cosmological particle production after inflation has been discussed for
time-dependent mass $m(t)$ in various situations.
Such particle production is called preheating\cite{Kofman:1997yn,
Greene:2000ew}, for which there is no serious need for careful
discussions about locality.
A serious discussion about locality is important when the system appears
to be static and a symmetry is related to it.
The most important topic in this section is how this relationship
between locality and symmetry is described in differential geometry.

It was first recognized in Ref.\cite{Enomoto:2020xlf} that the Fermion
preheating can be interpreted as the Landau-Zener
transition\cite{Zener:1932ws}, and the
idea has been extended in Ref.\cite{Enomoto:2021hfv, Enomoto:2022nuj} 
to solve cosmological problems of particle production.
Particle-antiparticle asymmetry is not discussed here, but when it is 
described by the multi-element Landau-Zener transition, there are seeds
of asymmetry in the interference between different kinds of the Stokes
phenomena\cite{Enomoto:2021hfv, Enomoto:2022nuj}.
The relationship between cosmological particle production and the
Landau-Zener transition is not discussed in detail in this paper, so
more details and further explanations are left to these papers.

\subsection{Constant electric field (constant curvature)}
Introducing a constant electric field in the $z$-direction, we
find
\begin{eqnarray}
\hbar\dot{u}_\pm&=&i(k+e E_0 t) u_{\mp}\mp i m u_\pm,
\end{eqnarray}
which can be written in the matrix form as
\begin{eqnarray}
i\hbar\frac{d}{dt}\left(
\begin{array}{c}
u_+\\
u_-
\end{array}
\right)&=&\left(
\begin{array}{cc}
m& -k-e E_0t \\
 -k-e E_0t & -m
\end{array}
\right)
\left(
\begin{array}{c}
u_+\\
u_-
\end{array}
\right).
\end{eqnarray}

We will try to improve the analytical perspective by starting with a
general formulation.
We first consider the (generalized) Landau-Zener
transition\cite{Zener:1932ws} with 
\begin{eqnarray}
\label{eq-simpleoriginalLZ}
i\hbar \frac{d}{dt}\left(
\begin{array}{c}
X\\
Y
\end{array}
\right)&=&\left(
\begin{array}{cc}
D(t) & \Delta(t)^*\\
\Delta(t) & -D(t)
\end{array}
\right)
\left(
\begin{array}{c}
X\\
Y
\end{array}
\right).
\end{eqnarray}
Decoupling the equations, we have\footnote{One might claim that the
equation can be solved immediately using special functions. 
The reason for the somewhat roundabout approach here is that we want to
examine the Stokes lines in the vicinity of the tangent space.
To understand the structure of the Stokes lines, we use 
the exact WKB developed in Refs.\cite{RPN:2017, 
Voros:1983, Delabaere:1993, Silverstone:2008, Pham:1988, CNP:1993,
DDP:1993, DDP:1997, EWKB:text, EWKB:text2, ExactWKB:textbook, Virtual:2015HKT}.}
\begin{eqnarray}
\ddot{X}-\frac{\dot{\Delta}^*}{\Delta^*}\dot{X}+
\left(-\frac{iD\dot{\Delta}^*}{\hbar\Delta^*}
+\frac{i\dot{D}}{\hbar}
+\frac{|\Delta|^2+D^2}{\hbar^2}
\right)X=0.\\
\ddot{Y}-\frac{\dot{\Delta}}{\Delta}\dot{Y}+
\left(\frac{iD\dot{\Delta}}{\hbar\Delta}
-\frac{i\dot{D}}{\hbar}
+\frac{|\Delta|^2+D^2}{\hbar^2}
\right)Y=0.
\end{eqnarray}
In the following only solutions of $X$ are examined.
To obtain equations similar to the Schrodinger equation, we introduce 
$\hat{X}$ defined by
\begin{eqnarray}
\label{eq-normalEWKBtrans}
\hat{X}&=&\exp\left(-\frac{1}{2}\int^t
	       \frac{\dot{\Delta}^*}{\Delta^*}dt\right)X.
\end{eqnarray}
For the decoupled equations, the equation for $\hat{X}$ is
\begin{eqnarray}
\ddot{\hat{X}}+\left(\frac{-iD\dot{\Delta}^*}{\hbar \Delta^*}
+\frac{i\dot{D}}{\hbar}+\frac{|\Delta|^2+D^2}{\hbar^2}
+\frac{\ddot{\Delta}^*}{2\Delta^*}-\frac{3(\dot{\Delta}^*)^2}{4(\Delta^*)^2}\right)\hat{X}&=&0,
\end{eqnarray}
which can be written as
\begin{eqnarray}
\hbar^2\ddot{\hat{X}}+\left(Q_0+\hbar Q_1+\hbar^2 Q_2\right)\hat{X}&=&0,
\end{eqnarray}
where
\begin{eqnarray}
Q_0&=& |\Delta|^2+D^2 \nonumber\\
Q_1&=& \frac{-iD\dot{\Delta}^*}{\Delta^*}+i\dot{D}\nonumber\\
Q_2&=& \frac{\ddot{\Delta}^*}{2\Delta^*}-\frac{3(\dot{\Delta}^*)^2}{4(\Delta^*)^2}.
\end{eqnarray}
Seeing the $\hbar$-dependence\footnote{Our assumption here is that
$\dot{\Delta}$ does not generate an additional factor of $\hbar$ or
$1/\hbar$ in the equation\cite{Enomoto:2020xlf, Enomoto:2021hfv}. }, the Stokes lines of the above equation
coincide with the simple equation\cite{EWKB:text, ExactWKB:textbook}
\begin{eqnarray}
\label{eq-triv}
\ddot{\hat{X}}&+&\frac{|\Delta|^2+D^2}{\hbar^2}\hat{X}=0,
\end{eqnarray}
where $V(t)\equiv -(|\Delta|^2+D^2)$ is called a ``potential'' of
the ``Schrodinger equation''.
When $\Delta(t)=\lambda t$ and $D(t)=D_0$, the above equation
is similar to a well-known problem of quantum mechanics (i.e, the
scattering by an inverted quadratic potential).
The solutions of $|\Delta(t)|^2+D(t)^2=0$ are called ``turning points''.
The main difference between the analysis using the exact WKB and
the conventional analysis is that the local analysis in the vicinity of the
Stokes line is mathematically well defined in the exact WKB.
This distinguishing property of the exact WKB is the most important when
considering local analysis.

\begin{figure}[ht]
\centering
\includegraphics[width=0.6\columnwidth]{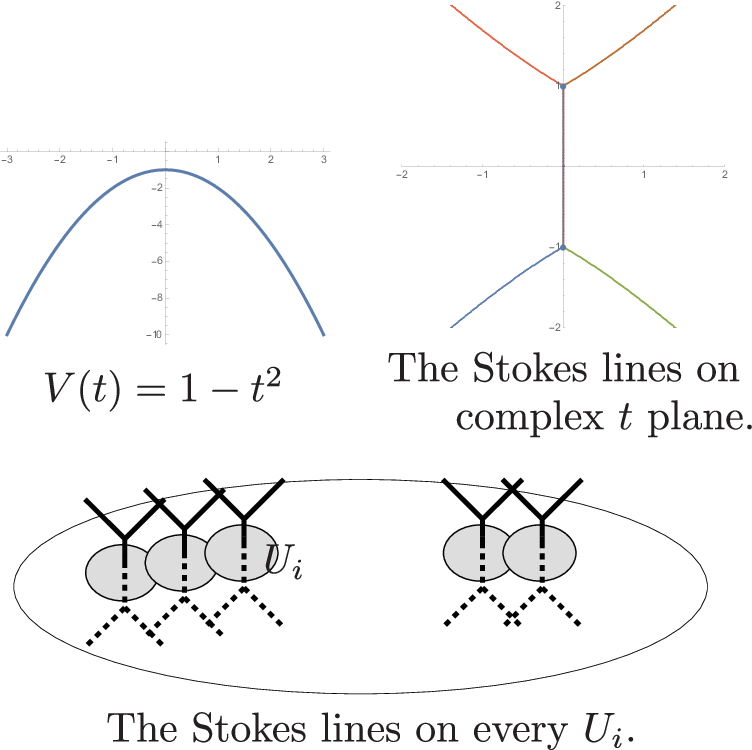}
 \caption{The typical ``potential'' of the equation is shown for
 $V(t)=1-t^2$.
The Stokes lines for the potential are shown on the complex $t$-plane.
They are presented on the top.
The Stokes lines appearing on $U_i$ of $M$ are illustrated in the bottom
 picture.
The top of the local ``potential'' must coincide with the contact point
 of the tangent space because of the local trivialization in each $U_i$.
This makes the Stokes lines coincide at the contact point and makes 
each $U_i$ looks completely the same.} 
\label{fig_fig1}
\end{figure}

There are a few things to be considered with care when using this
equation for solving the Schwinger effect.
The first and the most important is the definition of the local vacuum.
In the usual analysis of a constant electric field, one will find a
scattering problem by an quadratic potential, where the two vacuum states
(in and out states at 
$t=-\infty$ and $t=+\infty$ where the electric field is supposed to
disappear)
are defined as asymptotic
states.
Then, the Stokes phenomenon is assumed to occur between the two
vacuum states.
In this case, the gauge symmetry is used to explain the arbitrariness of
positioning the 
top of the potential hill.
However, as already explained, tangent spaces are naturally introduced
in manifolds, and these tangent spaces have the property for defining
the local vacuum states.
Therefore, instead of losing the locality of the analysis by assuming the vacuum
states at far away, we try to solve the problem by defining the vacuum in the
tangent space.
As we have already mentioned, a feature of the tangent space is that the
curvatur is zero in the space, and the connection must vanish at the
point of contact.
This is also a simple consequence of the local trivialization we have
described above.
This means that the gauge symmetry of each $U_i\in M$ must be used
carefully to make the connection vanish at the point.
Although it may not be trivial, one can see from the equation that this
makes the top of the quadratic potential coincide with the tangent 
space {\it in each $U_i$}, as is illustrated in Fig.\ref{fig_fig1}.
Then, they are laminated using gauge transformations.
The situation is quite similar to the moving frame in special relativity.
Similarly, since the tangent space is also a local inertial system, the
velocity (or $k$) is assumed to be negligible to account for the
inertial vacuum {\it seen by the generated particle}.
This changes the definition of the electric field $E_0$ in
the equation since the original $E_0$ was defined for an observer in
the laboratory.
Therefore, we denote the electric field in the equation described in the
vicinity of the tangent
space as $\hat{E}$ in the following calculations.
Now we have defined everything in the vicinity of the contact point of
the tangent space.
Note that the ``decoupling'' of the equations performed above cannot
be defined at the point of contact with the tangent space, as the
non-diagonal elements disappear from the matrix and the equations are
already decoupled there. 
This shows that there is no mixing in the tangent space by definition.
As discussed in the definition of manifolds, the Stokes phenomenon
should be considered on an open set $U_i$ defined in a neighborhood of
$p\in M$, and it is the physics around $p$ that is relevant for mixing solutions.
Therefore, the Stokes phenomenon is considered here in a neighborhood.
The above equations show that the real-time axis traverses the Stokes
line in the vicinity of the tangent space on every
$U_i$\cite{Matsuda:2023mzr}.
See also Fig.\ref{fig_fig1}.

Thus, when the local setting is made natural as a manifold, it can be seen
that the Stokes lines appear in the neighborhood
of the local tangent space without ambiguity of the time and the gauge.
This indicates that the Stokes Phenomenon is constantly mixing the
vacuum solutions.
Exactly speaking, the vacuum defined ``in'' the tangent space cannot
describe the mixing by definition, while 
the mixing is seen in the neighborhood of the tangent space placed on
each $U_i$.
This is the same situation as the asymptotic states.
In the same way as the correspondence with the vacuum is considered for
the asymptotic states, the correspondence with the vacuum solutions here is
considered in the vicinity of the tangent space.

Here, one might notice that the frame prepared for the generated
particles of the Schwinger effect actually represents an accelerated frame
called the moving frame. 
For this simple reason, the analysis of the Schwinger effect would not
be complete without an analysis of the Unruh effect.
This topic will be discussed in the next section.
In the remaining part of this section, we
are going to look at the time-dependent case in a little more detail to
show a simple example in which definition of the local vacuum is crucial
for the calculation.

\subsection{Time-dependent $E(t)$}
We will now try to show an example where it makes {\it no sense at all as
physics to find a naive global solution by considering extrapolation} to
an equation defined on an open set $U_i$.
It naturally depend on the situations whether a meaningful result can be
obtained by such extrapolation.
However, as far as we know, there has been no paper in which this point
is mentioned. 
In order to understand the problem without ambiguity,
consider the case where the electric field changes gradually and there
is no significant back-reaction from particle production.
If the electric field oscillates, our assumption here is that the time period
of the oscillation is much longer than the width of $U_i$.

\begin{figure}[ht]
\centering
\includegraphics[width=0.6\columnwidth]{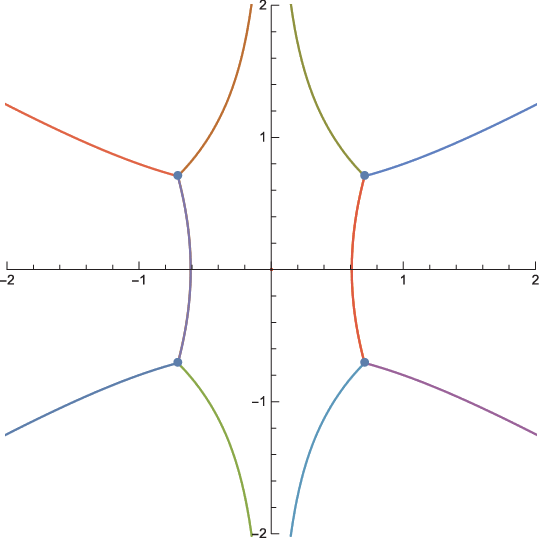}
 \caption{The typical Stokes lines of the scattering problem by a
 quartic potential are shown for $V(t)=1-t^4$. In contrast to the
 quadratic potential, the
 Stokes lines are not crossing the origin.}
\label{fig-quartic}
\end{figure}

Replacing the constant electric field $E_z(t)=E_0$ with a slowly varying
electric field $E_z(t)=\alpha t$, 
the problem becomes scattering by a ``quartic'' potential $V(t)\propto -t^4$.
The solution to this problem and the Stokes lines have been studied in
great detail by Voros\cite{Voros:1983}, and it has been found that the
Stokes lines act away from the origin.
See Fig.\ref{fig-quartic} for more details about the Stokes lines.
Details concerning conventional particle production can be found in
Ref.\cite{Enomoto:2020xlf}. 
What is important here is that unlike the case of the constant electric field, the gauge ambiguity
cannot naively shift the position of the quartic potential.
Also, although intuitively the probability of particle production should change
gradually if the electric field changes slowly, a conventional
calculation (scattering by the quartic potential solved for asymptotic
states) will not change in such a way.
Although it shoule be immediately obvious to anyone who looks at the equation,
we point out that the solutions of the equation, if
they are used for the asymptotic states, can not solve the problem at
all, even if the solutions are exact.
It is easy to finish the discussion by saying that the setting of the problem is bad, but here we
are going to try to clarify our understanding a little bit more.
For our purpuses, it is important to go back to the definition of the
manifold and solve the problem in local.

\begin{figure}[ht]
\centering
\includegraphics[width=0.6\columnwidth]{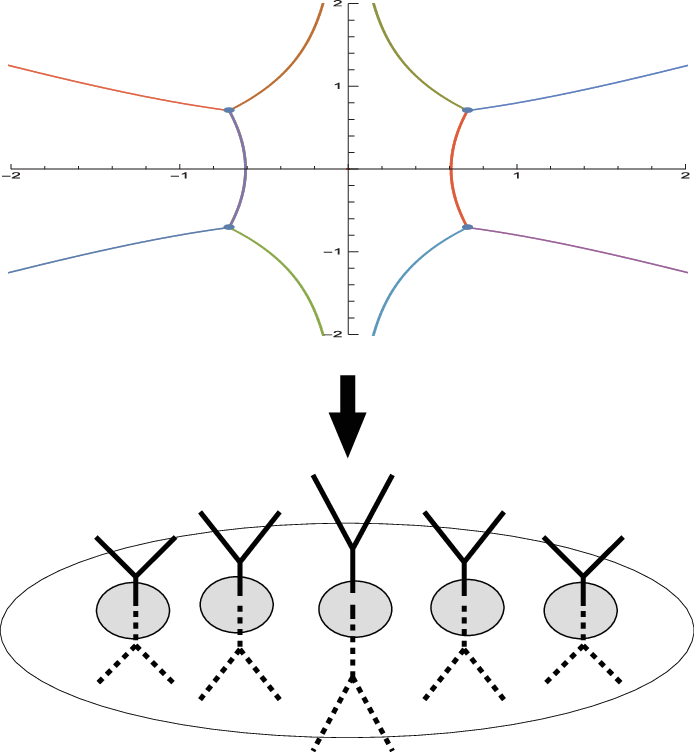}
 \caption{The upper picture shows the Stokes lines of the solution when
 the equation defined at $t=0$ is extrapolated to infinity. 
The lower picture shows the Stokes lines when equations are defined for
 each $U_i$ using local trivialization. 
One can see that the production rate is gradually changing with time (as
 the distance between two turning points are changing with time).}
\label{fig_discretized}
\end{figure}

Let us now rephrase the above question along the lines of the
manifold and the differential geometry.
First, consider an open set $U_i$ in the neighborhood of $t=t_p$
and let $A_z(t)=\frac{1}{2}\alpha t^2$ be expanded at $t=t_p$.
Then, we have the ``local potential'' defined for $U_i$ as
\begin{eqnarray}
V(t)|_{U_i}&\simeq& m^2-e^2\hat{E}_z(t_p)^2(t-t_p)^2,
\end{eqnarray}
where by using the gauge of $U_i$ the connection ($A_\mu$) is set to
zero at the contact point $t=t_p$.
$k\simeq 0$ is also expected for the inertial frame of the particle.
Here the tangent space and the local vacuum is defined at $t=t_p$.
This equation clearly meets the above requirement if the local vacuum is
defined in the tangent space.
See also Fig.\ref{fig_discretized}.
The particle production can be calculated locally on each $U_i\in M$ by
using $V(t)|_{U_i}$ and the
production rate gradually changes with time.
In this way, the generation rate can be calculated for the local time
($t_p$) on the local open set ($U_i$).
For cases like the one discussed here, the use of the
tangent space to define the local vacuum is {\it inevitable}.
The conventional calculation defining the asymptotic
vacuum at a distance could be convenient but not rigorous in mathematics.
We would like to stress the importance of understanding
benetits and risks of the definitions of the vacuum.
As far as we can understand, the serious risk of defining the vacuum in
the asymptotic state has not been explained in detail so far for the
Schwinger effect.

As we have already seen in the simple example of a slowly changing
electric field, it is very important and sometimes quite essential to
solve the field equations locally on the manifold by defining a local
vacuum in the tangent space.
Of course, when determining the {\it averaged} particle production rate in
the case of rapid oscillations or when the electric field is
sharply instantaneous, local evaluation of the particle
production rate should be useless for experimental observation. 
If the electric field changes rapidly, the field equation becomes a
simultaneous differential equation with the electromagnetic field
equation.
The Stokes phenomena in such cases can be extremely complex and require
appropriate approximations.\footnote{Analysis of these topics using the
Stokes lines can be 
found in Ref.\cite{Taya:2020dco}.
See also Refs.\cite{Nakazato:2021kcb,Kitamoto:2018htg}.}.
What we have highlighted in this section is the case in which local
calculation is crucial to obtain correct results while a ``naive'' 
extrapolation of the equation to the outside of the defined area
gives clearly a wrong answer. 
This distinction has never been explicitly recognized so far. 
We believe that the importance of the option of defining the {\it local}
vacuum in
the tangent space has been made very clear by the simple model described
in this section.

Let us now look at how such a definition of the vacuum has 
implications in relativity.
As mentioned above, particles created by the Schwinger effect can also
be affected by the Unruh effect.
In this case, the question is whether the two effects are independent
phenomena or whether we are simply seeing the same thing.
In the following, we present the solution to this problem using
differential geometry and the exact WKB.

\section{The Unruh effect and Hawking radiation on manifolds} 
\label{section-HWandU}
The previous section dealt with the case where there is no qualitative
ambiguity in 
defining the local vacuum, but in general relativity, even after
the connection (metric) is determined, there are still
further degrees of freedom left in the vierbein, which leaves ambiguity
in defining the local vacuum.
The most obvious difference is between the Lorenz frame and the local
inertial frame.
In mathematics, covariant derivatives are defined by using the
Lorenz frame, which gives a vierbein that is diagonal with respect to
the time-direction {\it in the neighborhood}.
On the other hand, in physics, the vacuum is defined for the local
inertial frame, which gives a vierbein that is diagonal with respect to
the time-direction {\it only at the contact point}.
In both cases, the tangent space is correctly defined at the point, but
there is a difference in physics in the neighborhood.

To understand the situation more clearly, let us start by introducing the notion of
the frame bundle in more detail.
Suppose that $\varphi_i(p)$ is the coordinate function $\{x^\mu (p)\}$
of $U_i$.
In the ``coordinate basis'', $T_p M$ is spanned by
$\{e_\mu\}=\{\partial/\partial x^\mu\}$, while the ``non-coordinate
bases'' is explained as
\begin{eqnarray}
\hat{e}_\alpha&=&e_\alpha^\mu\frac{\partial}{\partial x^\mu},\,\,\,\,
e_\alpha^\mu\in GL(m,\mathbb{R}),
\end{eqnarray}
where the coefficients $e_\alpha^\mu$ are called vierbeins.
Since $U_i$ is homeomorphic to an open subset $\varphi(U_i)$ of
$\mathbb{R}^m$ and each $T_pM$ is homeomorphic to $\mathbb{R}^m$,
$TU_i\equiv \bigcup_{p\in U_i}T_pM$ is a $2m$-dimensional manifold,
which can always be decomposed into a direct product $U_i\times
\mathbb{R}^m$.
This means that the local theory at that point (not in the
neighborhood) is nothing but special relativity. 
Note that in differential geometry everything starts with local trivialization.
Given a principal fibre bundle $P(M,G)$, one can define an associated
fibre bundle as follows.\footnote{The explanation here is in the
opposite direction to the description of the principal bundle we have already
given. Previously, we started from the fibre bundle to reach at the notion of
the principal bundle.
The explanation here is useful when the structure group is determined first.}
For $G$ acting on a manifold $F$ on the left, one can define an action
of $g\in G$ on $P\times F$ by
\begin{eqnarray}
(u,f)&\rightarrow&(ug,g^{-1} f)
\end{eqnarray}
where $u\in P$ and $f\in F$.
Now the associated fibre bundle is an equivalence class $P\times F/G$ in
which $(u,f)$ and $(ug,g^{-1} f)$ are identified.
For a point $u\in TU_i$, one can systematically decompose the
information of $u$ into $p\in M$ and $V\in T_pM$.
As we have mentioned, this leads to the projection $\pi$ :
$TU_i\rightarrow U_i$.
Normally, $\hat{e}_\alpha$ is requested to be orthonormal with respect
to g;
\begin{eqnarray}
\mathrm{g}(\hat{e}_\alpha,\hat{e}_\beta)=e_\alpha^\mu e_\beta^\nu
 \mathrm{g}_{\mu\nu}=\delta_{\alpha\beta},
\end{eqnarray}
where $\delta_{\alpha\beta}$ is replaced by $\eta_{\alpha\beta}$ for
the Lorentzian manifold.
The metric is obtained by reversing the equation
\begin{eqnarray}
 \mathrm{g}_{\mu\nu}=e^\alpha_\mu e^\beta_\nu\delta_{\alpha\beta}.
\end{eqnarray}
What is important for our discussion is that
in an $m$-dimensional Riemannian manifold, the metric tensor
$\mathrm{g}_{\mu\nu}$ has $m(m+1)/2$ degrees of freedom while the vielbein
has $m^2$ degrees of freedom.
For $m=4$, we have $10$ for the metric while $16$ for the vielbein.
They are not identical.
Each of the bases can be related to the other by the local orthogonal
rotation $SO(m)$, while for the Lorentzian manifold it becomes $SO(m-1,1)$.
The dimension of these Lie groups is given by the difference between the degrees
of freedom of the vielbein and the metric.
This shortly means that there are many (uncountable) choices for
non-coordinate bases even after the metric is identified.
This point will be very important when one looks at the Unruh
effect\cite{Fulling:1972md, Davies:1974th, Unruh:1976db}. 
The local inertial frame and the Lorentz frame have the same metric and
are defining the same tangent space at the point.
However, they are distinguished by the vierbein.
The difference in the vierbein is essential in the search for the Stokes
phenomenon of the Unruh effect\cite{Matsuda:2023mzr}.

We describe the frame bundle further below.
Associated with a tangent bundle $TM$ over $M$ is a principal bundle
called the frame bundle $LM\equiv \bigcup_{p\in M}L_p M$
where $L_p M$ is the set of frames at $p$.
Here we have a natural coordinate basis 
$\{\partial/\partial x^\mu\}$ and a ``frame''
 $u=\{X_1,...,X_m\}$ at $p$ is expressed by the non-coordinate basis
\begin{eqnarray}
X_\alpha&=&X^\mu_\alpha \left.\frac{\partial}{\partial x^\mu}\right|_p
\end{eqnarray}
 where $(X^\mu_\alpha)\in GL(m,\mathbb{R})$.   
If $\{X_\alpha\}$ is normalized by introducing a metric, the matrix
$(X_\alpha^\mu)$ becomes the vielbein.

The following point is very important for our discussion.
A natural coordinate basis is prepared for $U_i\in M$
and {\it the inertial system is defined using a non-coordinate basis in
the tangent space.}
This procedure naturally gives a notion of the ``moving
frame''\cite{Misner:textbook}, as the inertial frame seems to be rotated
from time to time by the Lorentz transitions of the vielbeins, somewhat like
spinning tea cups in amusement parks.\footnote{One can see more explanations in
Ref.\cite{Misner:textbook}, in which figures for the moving frame can be
found.} 
The Thomas precession is explained by the fact that multiple Lorentz
transformations with different directions produce a rotation of the
intrinsic space of the observer.
If the observer stays in the same frame (no acceleration), there is
no motion in the direction of the fibre of the frame bundle.
In this case, the connection vanishes by definition.
Therefore, for an inertial observer, the distinction
between coordinate and non-coordinate systems will be quite ambiguous.
However, if one wants to describe an observer in accelerated
motion on flat space-time, one has to define the local and subjective 
inertial frame
for the observer in the tangent space, which moves on the frame bundle
in a non-trivial way. 
This defines the section of the observer on the frame bundle.
This means that on this section, one has to
laminate the bundle by using non-trivial coordinate transformations.
The Thomas precession is explained that the lamination by the Lorentz
transformation causes rotation of the intrinsic space.
This (the observer's non-trivial section of the frame bundle) 
introduces the connection {\it for the observer}, although the
space-time is flat.

To be more specific, the vielbeins for constant acceleration ($a$) 
in the two-dimensional space-time  at $\tau=\tau_A$ is
\begin{eqnarray}
\label{eq-vier-const}
(e_A)_\alpha^\mu&=&\left(
\begin{array}{cc}
\cosh a(\tau -\tau_A)&\sinh a(\tau -\tau_A)\\
\sinh a(\tau -\tau_A)&\cosh a(\tau -\tau_A)
\end{array}
\right).
\end{eqnarray}
Indeed, for such constant acceleration, the vielbeins have completely the
same form for any time $(\tau)$.
At $\tau=\tau_B$, we have 
\begin{eqnarray}
(e_B)_\alpha^\mu&=&\left(
\begin{array}{cc}
\cosh a(\tau -\tau_B)&\sinh a(\tau -\tau_B)\\
\sinh a(\tau -\tau_B)&\cosh a(\tau -\tau_B)
\end{array}
\right).
\end{eqnarray}
The transformation $(e_A)_\alpha^\mu\rightarrow (e_B)_\alpha^\mu$ on the
frame bundle is the Lorentz transformation;
\begin{eqnarray}
L_{AB}&=&\left(
\begin{array}{cc}
\cosh a(\tau_B -\tau_A)&-\sinh a(\tau_B -\tau_A)\\
-\sinh a(\tau_B -\tau_A)&\cosh a(\tau_B -\tau_A)
\end{array}
\right).
\end{eqnarray}
Note that the situation is very similar to the Schwinger effect for a
constant electric field.
In the Schwinger effect, the potential has been shifted by gauge
transformation and the equation looks always the same on each $U_i$.
In the Unruh effect, the Lorentz transformation gives
vielbeins of exactly the same shape on each $U_i$.
In both cases, the observer is always looking at the same physics on a
static manifold.

Our primary question here is ``Is the same local analysis possible also
for the Unruh effect, if we follow the previous calculations of the
Schwinger effect?''
Our answer is ``No''.
What is important here is that the mathematical definition of the covariant
derivative uses the Lorenz frame in which the vierbein is diagonalized 
{\it in the neighborhood}.
On the other hand, as is shown explicitly above, the vierbein of the
local inertial frame is diagonalized {\it only at the point}.
(Note that the off-diagonal elements vanish at the point since $\sinh
a(\tau-\tau_A)=0$ at $\tau=\tau_A$.) 
Since the covariant derivatives are defined for the Lorentz frame while
the Unruh effect is defined for the inertial frame, 
it seems impossible to examine the Stokes phenomena of the Unruh
effect directly (and locally) in terms of the field
equations\cite{Matsuda:2023mzr}. 
This mismatch has prevented the local analysis of the Unruh effect for a long time.

Now we consider the physics that observers see in the Unruh effect.
As far as the acceleration seen by the observer is constant, 
the physics seen by the observer is indistinguishable at any time due to
the equivalence classes defined for the manifold.
(More particularly, the observer feels the same vierbein all the time.)
This is purely a mathematical consequence.
If a dynamical effect (particle production) is manifested in such a
situation, it must be explained by the vierbein of the local inertial
frame.
Extrapolating the coordinates to infinity and considering a global map
as the Bogoliubov transformation is 
conceptually {\it unacceptable}, even if it could be commonly used as a
method\cite{Birrell:textbook}. 
In fact, it is known that such methods can lead to unnatural
entanglements appearing between regions that should be uncorrelated.
To argue the legitimacy of our local computation on
manifolds, we must address this issue in this paper.

First of all, consider what happens if the scalar field equations
were written down for an accelerating observer.
Rindler coordinates are a specific set of coordinates used in the
context of special relativity to describe the motion of an observer
undergoing constant proper acceleration in flat spacetime. 
Thus, the Rindler coordinates form {\it a coordinate chart (not a
frame)} that covers a 
specific region of Minkowski spacetime known as the Rindler wedge. 
Following the concept of manifolds, the simplest local inertial system
is defined as a tangent space in the neighborhood of the point with
zero velocity in Rindler's coordinates.
 If one starts with the conventional Rindler
metric\cite{Birrell:textbook,Misner:textbook,
deGill:2010nb};\footnote{Note that we do not use the metric
in our calculations. 
The reason for showing a metric here is just to explain that the Stokes
phenomenon cannot be calculated by the metric.}
\begin{eqnarray}
ds^2&=&-\left(1+a x\right)^2 dt^2+dx^2,
\end{eqnarray}
 one will not be able to find the required Stokes lines.
As is immediately apparent from the metric (and the equations), the
field equation  
does not yield the local Stokes phenomenon as was the case
with the Schwinger effect\cite{Enomoto:2022mti}.
This indicates that there may be a fundamental error in the way of
setting the problem.
Here, we shall stick to the locality of the issue.

\begin{figure}[ht]
\centering
\includegraphics[width=0.8\columnwidth]{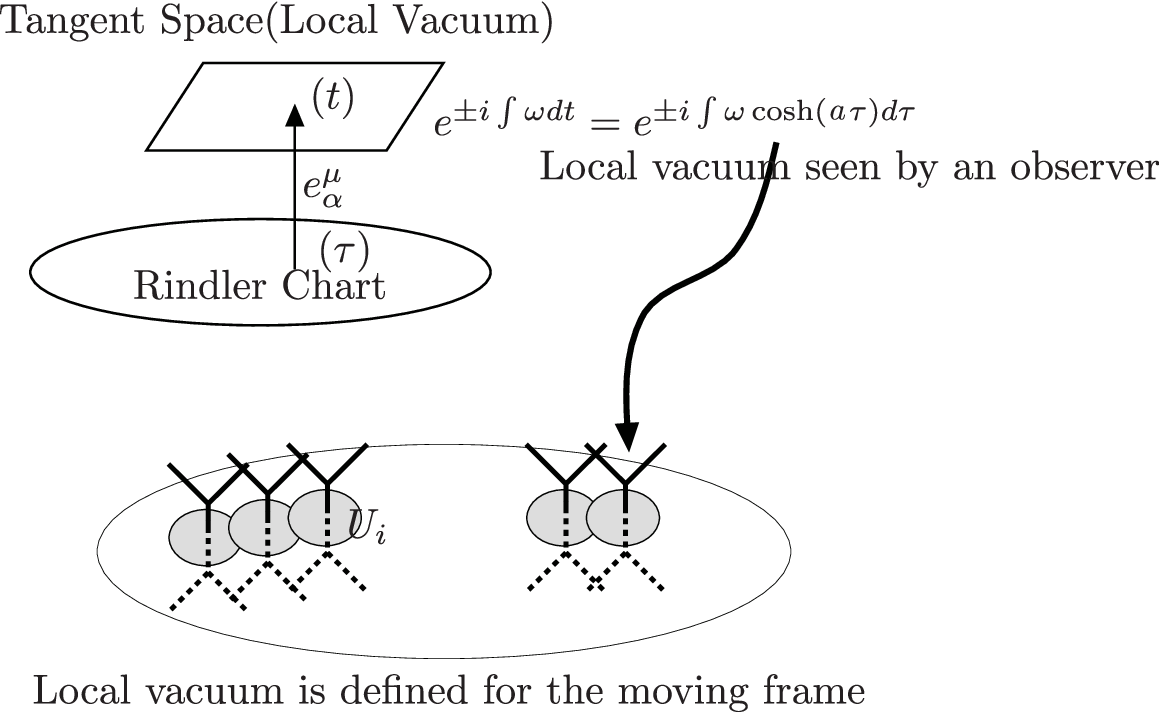}
 \caption{The local vacuum defined in the tangent space is seen by an
 accelerating observer using the vierbein.
The Stokes phenomenon on the Moving frame is illustrated in the bottom.
The situation is always the same for the observer if the acceleration is
 constant.
The observer sees always the same Stokes phenomenon on the moving frame,
where the subjective vacuum for the observer is defined in the tangent
 space using the inertial frame of the observer.} 
\label{fig-Unruh}
\end{figure}

Since the metric (covariant derivatives) does not explain the local Stokes phenomenon, the only
way left for us is to use the vierbein.
The problem is that even if the vierbein is used for the calculation,
the same trivial result is obtained once the field equations are written
down, as the Lorentz frame and the inertial frame are not
distinguishable by the metric.
Therefore, the vierbein must be used without going through the field
equations. 
Now consider what an accelerating observer would see if the observer
 looked directly at the vacuum (vacuum solutions) 
defined in the observer's local inertial space.
The vacuum is observed here by the particle itself, which is
ejected from ``the vacuum''.\footnote{Alternatively, one can introduce
the Unruh-DeWitt detector as an observer. The problem has been solved in
Ref.\cite{Matsuda:2024ydu} by paying serious attention to the locality
and the differential geometry. The result is consistent with this paper.}
Since such particles have no momentum in their unique frame, we can
neglect the $x$-dependent component of the vacuum solution.
Considering $dt= \cosh\left(a \tau\right) d\tau$ from
Eq.(\ref{eq-vier-const}), we have for the time-dependent part;
\begin{eqnarray}
\label{eq-stokes1}
e^{\pm i \int \omega dt}&=&
e^{\pm i \int \omega \cosh(a \tau) d\tau},
\end{eqnarray}
where $dt$ is needed for the calculation, even if $\omega$ is a
constant.
The situation is illustrated in Fig.\ref{fig-Unruh}. 
We have already analyzed the Stokes phenomenon of the above function in
detail in Ref.\cite{Enomoto:2022mti, Matsuda:2024ydu}.
As the details of the Stokes phenomenon of the solution are not the
issue of this paper, we will present only the results here.
See Ref.\cite{Enomoto:2022mti,Matsuda:2024ydu} if the reader is
interested in the details about the Stokes phenomenon.
One thing that should be noted is that the
result does not meet the conventional (global) calculation by a factor
of two.
This point has to be explained here.

To make the point clear, we consider the entanglements that appear in
conventional calculations\cite{Birrell:textbook}.
 In the conventional calculation, the factor of 2 is added
when taking a trace on the entanglements in the distant wedge.
This is because the particle creation in the right wedge is always
accompanied with that in the left wedge as 
$e^{-\pi \omega /a} b_{\bm{\mathrm{k}}}^{L \dagger} b_{\bm{\mathrm{k}}}^{R \dagger}
|0_L\rangle \otimes 
|0_R\rangle$,
where $b_{\bm{\mathrm{k}}}^{L \dagger}$ and $b_{\bm{\mathrm{k}}}^{R \dagger}$ are the creation
operators 
in the left and the right wedges, respectively.
One can see that there is a duplication of the factor, which can be
separated as
$\left(e^{-\pi \omega /2a} b_{\bm{\mathrm{k}}}^{L \dagger} \right)
\left(e^{-\pi \omega /2a} b_{\bm{\mathrm{k}}}^{R \dagger}\right)  |0_L\rangle \otimes
|0_R\rangle$.
This duplication does not appear if the calculation is local.
The entanglement appeared because of the extrapolation beyond the
applicable range of the moving frame that defines the observer's
vacuum.
Also, to be more rigorous, defining a vacuum in the chart is not a
preferred method.
In our calculation, we have the factor $e^{-\pi \omega /2a}$
because it is derived from the {\it local} Stokes phenomenon.
For a local pair creation, we will find $e^{-\pi \omega /a}$ but in our
case both particles are observed and there is no trace-out.
If we {\it assume} the entanglement between the distant wedge, our
result can reproduce the usual Unruh effect calculations, but since our
local calculation is rigorous as mathematics, there is no need to 
assume such entanglement with a distant wedge.
This is the Bogoliubov coefficient of the state mixing in the vacuum
when the vacuum is seen by an accelerating observer.
Let us be more precise.
Noting that the vierbein connects the inertial frame (vacuum) and the
observer, we rewrite $\phi$ in the vacuum as
\begin{eqnarray}
\phi(t,{\bm{\mathrm{x}}})&=&\sum_{\bm{\mathrm{k}}} e^{-i {\bm{\mathrm{k}}}\cdot
{\bm{\mathrm{x}}}}\left[
a_{\bm{\mathrm{k}}} e^{-i\int \omega dt}
+a_{-\bm{\mathrm{k}}}^\dagger e^{+i\int \omega dt}
\right],
\end{eqnarray}
where $\omega$ is constant but $dt$ is required to introduce
the vierbein in an explicit form.
If a vierbein gives $dt=(e(\tau))^t_\tau d\tau$, mixing
of the solutions (when it is seen by an accelerating observer whose
proper time is $\tau$) 
after crossing the Stokes line is written by
\begin{eqnarray}
e^{\pm i\omega \int^\tau   (e(\tau'))^t_\tau d\tau'}&\rightarrow&\alpha_\pm
e^{\pm i\omega\int^\tau  (e(\tau'))^t_\tau d\tau'} 
+\beta_\pm e^{\mp i\omega \int^\tau
(e(\tau'))^t_\tau d\tau'}.
\end{eqnarray}
Then, we have 
\begin{eqnarray}
\phi(t,{\bm{\mathrm{x}}})&=&\sum_{\bm{\mathrm{k}}} 
e^{-i {\bm{\mathrm{k}}}\cdot
{\bm{\mathrm{x}}}}\left[
a_{\bm{\mathrm{k}}} 
\left(\alpha_- e^{- i\omega\int^\tau  (e(\tau'))^t_\tau d\tau'} 
+\beta_- e^{+ i\omega \int^\tau
(e(\tau'))^t_\tau d\tau'}
\right)
\right.\nonumber\\
&&\left.+a_{-\bm{\mathrm{k}}}^\dagger 
\left(
\alpha_+
e^{i\omega\int^\tau  (e(\tau'))^t_\tau d\tau'} 
+\beta_+ e^{- i\omega \int^\tau
(e(\tau'))^t_\tau d\tau'}\right)
\right],
\end{eqnarray}
which suggests that the particle production is seen by $\beta_\pm\ne 0$.
We call this factor the Bogoliubov coefficient, and 
this factor should be considered as the Boltzmann factor of the Unruh effect.
With entanglement, the probability ($P_1$ for a particle production in
our calculation) is
squared because there are two particles produced {\it but only one particle
is observed.}
This gives $P_{entangled}=P_1^2$ in our calculation. 
In this way, our calculations will give the same
result as the conventional calculation {\it if the entanglement is
included by hand}.
In short, if we ``assume'' entanglement, our calculation coincides
with the conventional calculation, but obviously our local calculation
does not require the entanglement. 
In our previous paper\cite{Matsuda:2024ydu}, we have calculated the
Boltzmann factor of the 
Unruh-DeWitt detector and found the same result.
Therefore, our local calculations are consistent
between the Unruh effect and the Unruh-DeWitt detector, but are not
consistent with conventional calculations.
The discrepancy is explained by the entanglement.
Our local analyses in terms of the differential geometry and the exact
WKB allow calculating the probability of single
particle observation as the single particle production.
Then, our results differ by a factor of two from the
usual Unruh Effect and the Unruh-DeWitt detector calculations.
So what happens if we apply our calculations to Hawking radiation?
Conventional calculations of the Unruh effect treat infinite inertial space
as real, so the existence of such spaces cannot be
assumed for Hawking radiation. 
On the other hand, our calculations are always locally defined, so the
same calculation of the Unruh effect can be performed in the vicinity of 
the black hole horizon.
When a pair of particles is created in the vicinity of the horizon, one
particle inside the horizon can have negative energy (when it is seen from
the outside), and the other in the outside can have positive energy.
Again, as there are two particles produced but only one particle is
observed as radiation, the total production probability for a particle
radiation  
is squared to obtain results consistent with Hawking
radiation\cite{Enomoto:2022mti, Matsuda:2023mzr}.
In this case, unlike the Unruh effect, we do not have to assume unnatural entanglement at a distance.
Thus, if all calculations are performed faithfully to mathematics of the
differential geometry,
strong doubts will arise about the entanglement of the 
Unruh effect.\footnote{We are not claiming that our analysis is able
to provide a proof that there is no entanglement in the conventional
global calculations. 
It is an indisputable fact that entanglement appears in the global
calculation, although the mathematical consistency is not
reliable because of the extrapolation. }

For completeness, we will now consider the Unruh effect as the
acceleration changes, which is similar to the slowly changing electric
field in the Schwinger effect.
Simply because the general calculations of the Unruh-DeWitt detector are
somewhat different from the Unruh effect itself, we will
treat them as different.
See Ref.\cite{Matsuda:2024ydu} for more details about what the local 
calculations of the Unruh-DeWitt detector look like if it is defined
locally using the differential geometry and the exact WKB.
Note also that the Unruh-DeWitt detector requires an explicit
interaction with the detector to be included, which is unlikely to be
applicable directly to Hawking radiation.

\subsection{When the acceleration rate is slowly time-dependent}
Let us first derive the Rindler coordinate
for the case of constant acceleration using a somewhat roundabout
approach.
For simplicity, we restrict the motion to the x-axis direction.
If the acceleration seen by the inertial system is $a(t)$ and the
acceleration seen by an observer moving at the speed of $v(t)$ with
respect to the inertial system is $a'(t')$, then the following relationship holds.
\begin{eqnarray}
a'&=&\left(1-\frac{v^2}{c^2}\right)^{-\frac{3}{2}}a.
\end{eqnarray}
Since the acceleration seen by an observer in accelerated motion with
respect to an inertial system is $a'$, $a'$ is a constant in the
conventional Unruh effect.
After integrating both sides with respect to the time $t$, we find 
\begin{eqnarray}
a_0t&=&\frac{v}{\left(1-\frac{v^2}{c^2}\right)^{\frac{1}{2}}},
\end{eqnarray}
where $a(t)=dv/dt$ and $v(0)=0$ has been used.
This ($v(0)=0$) means that the contact point with the tangent space is placed at $t=0$.
Solving the above equation for $v(t)$, we find
\begin{eqnarray}
v(t)&=&\frac{a_0t}{\sqrt{1+\left(\frac{a_0t}{c}\right)^2}},
\end{eqnarray}
which can be used to calculate the relation between the time coordinates
as 
\begin{eqnarray}
T'&=&\int^T_0\sqrt{1-\frac{v(t)^2}{c^2}}dt\nonumber\\
&=&\int^T_0\frac{1}{\sqrt{1+\left(\frac{a_0t}{c}\right)^2}}dt\nonumber\\
&=&\frac{c}{a_0}\sinh^{-1}\left(\frac{a_0}{c}T\right).
\end{eqnarray}
Finally, we find
\begin{eqnarray}
\frac{a_0}{c}T&=&\sinh \left(\frac{a_0}{c}T'\right).
\end{eqnarray}
Since the function $T$ of $T'$ (i.e, $T(T')$) is periodic in the
direction of the imaginary 
axis of $T'$, we can expect any vacuum function $F(T)$ described by the observer's
time $T'$ to be periodic in the observer's complex time.
Here a simple question would arise.
The periodic function for imaginary $T'$ seen here is merely a
parameterization of the elliptic function so that it takes an infinite
limit for one of its double periodicity.
The simple answer is that this may be a very special case due to the
assumption that the acceleration is a constant.
Let us see this point in more detail.
In the above calculation, we simply had
\begin{eqnarray}
\int a' dt&=&a_0 t +C,
\end{eqnarray}
where we set $C=0$ by $v(0)=0$.
Let us relax the condition of this calculation and try to examine the
following;
\begin{eqnarray}
\int a' dt&=&f(t).
\end{eqnarray}
Then, for $f(t)=\alpha t + \beta t^2$ we have
\begin{eqnarray}
T'&=&\int^T_0\frac{c^2}{\sqrt{c^2+(\alpha t+\beta t^2)^2}}dt,
\end{eqnarray}
which gives an elliptic integral after Legendre's 
transformation and thus $T$ as the function of $T'$ is described by an
elliptic function. 
Obviously, if we consider $\beta\ne 0$, the periodicity of the function
$T(T')$ is not a simple imaginary.
The above ``coordinate system'' for the slowly varying acceleration do
not have the special properties of the Rindler coordinates.
More specifically, the situation cannot be always
the same for the observer.
This means that unlike the conventional Unruh effect for a constant
acceleration, the vierbeins cannot be moved to the same 
form by the Lorentz transformation.
This situation is quite similar to the case which appeared when we have
considered the weakly time-dependent electric field in the Schwinger effect.
Note that our local analysis only considers slices of the local elliptic
function at
the real axis and do not extrapolate it to infinity.
The most familiar example of the elliptic function is probably the motion of a
pendulum.
Indeed, if we set
\begin{eqnarray}
f(t)&=&\int a_0 \cos \omega t\nonumber\\
&=& \frac{a_0}{\omega} \sin \omega t\simeq a_0 t
\end{eqnarray}
 for an ``oscillation'',\footnote{This ``oscillation'' is not defined 
for the observer's time $t'$.} we can see that the approximate
solution at $t=0$ is obtained for the Unruh effect in the similar way as
a motion of a pendulum.\footnote{Jacobi functions are complex-valued functions
of a complex variable $z$ and a parameter $m(=k^2)$. Using the elliptic
integral of the first kind $K(m)$, the Jacobi $sn$ function has two
periods $4K(m)$ and $2K(1-m) i$. In the present case we have $m<0$,
while for an pendulum it becomes $m>0$.} 
We comment on the case of solving it at other times ($t\ne0$).
If we consider the Unruh effect at $t=t_0$, the inertial condition is now
$v(t_0)=0$. 
We thus have 
\begin{eqnarray}
f(t)&=&\int a_0 \cos \omega t\nonumber\\
&=& \frac{a_0}{\omega} \sin \omega t+C\nonumber\\
&\simeq& \left.\frac{a_0}{\omega} \sin \omega t\right|_{t=t_0}
+\left.\left(\frac{a_0}{\omega} \sin \omega t \right)'\right|_{t=t_0}
(t-t_0)+C ,
\end{eqnarray}
where the inertial condition gives $C=\frac{a_0}{\omega} \sin \omega t_0$.
Finally, we have
\begin{eqnarray}
f(t)&\simeq&a'(t_0) (t-t_0)
\end{eqnarray}
for $a'(t)=a\cos \omega t$.\footnote{Here the primes are used in two
different ways. The observer's $a'$ and $T'$ should be distinguished
from the derivatives.}
Using this result and the previous calculation for deriving $T(T')$, one
can find the Stokes phenomenon on the local space 
($U_i$).
These analyses explain how the periodicity of the Unruh effect in the
imaginary $T'$ direction is distorted by
the time-dependent acceleration and how the Unruh effect can be
calculated on a local space of the manifold without extrapolating the
space to infinity.

\section{Conclusions and Discussions}
\label{sec-concdis}
\hspace*{\parindent}
In this paper, the relationship between the Schwinger and Unruh effects
has been discussed on the basis of their similarities as
theories on manifolds.
The two phenomena, which at first sight appear to be the same, turn out
to be caused by completely different sources when one looks at the local
structure of the manifold.
As the idea of defining the vacuum in the local tangent space could be
unfamiliar in the analyses of the Schwinger effect and the Unruh effect,
this part of the article was explained in particular detail.
In this paper, the case of a gradual change in curvature has been
 considered as an example where the correct answer can only be obtained
 when the vacuum is defined in the {\it local} tangent space.
In our local analysis, the entanglement of the Unruh effect appears to
be an apparent one due to the extrapolation of the coordinate system.
On the other hand, our result agrees with Hawking radiation, where the
basic procedures of differential geometry can be used.

We will summarise the reasons why the factor of 2 difference
appeared between the conventional calculation of the Unruh effect and ours.
As we have described in detail in Sec.\ref{section-HWandU}, the direct reason
for this difference is the entanglement between distant wedges.
This entanglement does not actually exist.
The problem arose because of the following problems of the conventional
calculation.
\begin{itemize}
\item The moving frame was not moving. This ruins the mathematical
      consistency of the calculation.
\item The Rindler ``vacuum'' was defined in the chart, but as we have
      described in detail in Sec.\ref{sec-diff}, the chart is
      not a good place for defining a vacuum.
      One must remember how Lie algebra is introduced in differential
      geometry.
\item The Rindler chart describes eternal acceleration, but the observer
      cannot be accelerated eternally in a flat spacetime.
      Mathematically, there is no problem at this point, as the chart
      can be chosen in any way.
      The problem arises when one ignores mathematical consistency and
      behaves as if the chart is ``real''.
\item It must be remembered that the ``chart and frame'' cannot
      be used outside the application area.
      When dealing with an accelerating observer, either frame or chart
      has to be Lorentz-transformed. 
      Normally the frame is Lorentz-transformed and the frame is called
      the moving frame.
      Otherwise the mathematical consistency will be lost outside the
      application area\cite{Misner:textbook}.
\item The most serious problem with defining a vacuum in the chart appears
      when information about the far future or the far distance
      (arbitrarily defined by the people who have chosen the
      chart) is used for the calculation.
\item In the conventional calculation, the quantum states are assumed to be
      coherent over the whole space-time on the Rindler chart.
      The information in the far distance is used because of the coherence.
      The entanglement of the Unruh effect appears in this way.
      Of course, under normal circumstances, it would be inconceivable
      to define a state that is coherent to infinity on a chart that is
      merely a landmark.
\item If the calculation is faithful to differential geometry and no
      ``assumption'' is made in physics, the information to be used in
      physics is normally in the frame and the 
      vierbeins, not in the chart.
\end{itemize}
Despite its critical importance in this field, differential geometry has
not been used in good faith in the analysis of the Unruh effect.
The reason is that the procedures of the conventional calculation
cannot agree with the basic concepts of differential geometry.

If one attempts to solve the problem faithfully to the concepts of
differential geometry, one will find that tools are needed to derive the
Unruh effect locally.
The fact that such tools were not discovered has postponed the issue of
the Unruh effect.
The discovery of the local analysis by the exact WKB is a breakthrough in
this sense.

Now we would like to rephrase our most important conclusion.
``The entanglement between distant wedges of the Unruh effect is an
illusion caused by a computational glitch.''

We hope that our local analyses presented in this paper will help people
understand more clearly the physics of the quantum field theory on curved
manifolds.

\section{Note added in proof}
Einstein had concerns about entanglement and the locality of field
theory and relativity, and this is known as ``Einstein's concern''.
Einstein recognized this problem early on and presented its definition
and solution in his paper known as EPR\cite{EPR-paper}.
Later, experiments showed that the local realism presented in the EPR
paper was wrong, and this issue remains unresolved to this day (or one
might say that this remains to be a concern). 
This problem is also known as the ``quantum twin problem'', and the
results of the experiments are expected to be announced in the near
future. 
The concern relates to field theory in general, and the answer is still
unknown.

As the reader probably realizes, the most serious concern is the
entanglement that occurs between wedges that are already isolated at the
time of particle generation.
On the other hand, our calculations discuss the physics at the moment
when particles are generated locally, so at least with regard to the
particle generation mechanism, we can discuss this without touching on
the subsequent problems of entangled particles. 
The situation in our paper is the same as in normal field theory.
While our paper does not address Einstein's concern, it does provide
the first calculation faithful to the locality of field theory with
regard to the derivation of the Unruh effect.

\end{document}